\documentclass[times]{article}

\usepackage[cmex10]{amsmath}
\usepackage{amssymb}
\usepackage{amsthm}
\usepackage{graphicx,psfrag,epsf}
\usepackage{enumerate}
\usepackage[round,authoryear]{natbib}
\usepackage{clrscode3e}
\usepackage{url}
\usepackage{float}
\usepackage{todonotes}
\usepackage{xr}

\graphicspath{{figs/}}

\floatstyle{ruled}
\newfloat{code}{tpbh}{aux}
\floatname{code}{Code}

\newcommand{\Foreach}{\kw{for each}\,\Indentmore}
\newcommand{\Requires}{\kw{requires}\,\Indentmore}
\newcommand{\Ensures}{\kw{ensures}\,\Indentmore}

\renewcommand{\For}{\kw{for}\,\Indentmore}
\renewcommand{\While}{\kw{while}\,\Indentmore}

\newcommand{\Atomic}{\kw{atomic}\,}

\begin{document}

\title{Parallel resampling in the particle filter}
\author{Lawrence M. Murray\thanks{L.M. Murray (corresponding author) is with the Department of Statistics, University of Oxford.}, Anthony Lee\thanks{A. Lee is with the Department of Statistics, University of Warwick.} and
  Pierre E. Jacob\thanks{P.E. Jacob is with the Department of Statistics,
    University of Oxford.}}
  \maketitle

\begin{abstract}
Modern parallel computing devices, such as the graphics processing
unit (GPU), have gained significant traction in scientific and
statistical computing. They are particularly well-suited to
data-parallel algorithms such as the particle filter, or more
generally Sequential Monte Carlo (SMC), which are increasingly used in
statistical inference. SMC methods carry a set of weighted
\emph{particles} through repeated propagation, weighting and
resampling steps. The propagation and weighting steps are
straightforward to parallelise, as they require only independent
operations on each particle. The resampling step is more difficult, as
standard schemes require a collective operation, such as a sum, across
particle weights. Focusing on this resampling step, we analyse two
alternative schemes that do not involve a collective operation
(Metropolis and rejection resamplers), and compare them to standard
schemes (multinomial, stratified and systematic resamplers). We find
that, in certain circumstances, the alternative resamplers can perform
significantly faster on a GPU, and to a lesser extent on a CPU, than
the standard approaches. Moreover, in single precision, the standard
approaches are numerically biased for upwards of hundreds of thousands
of particles, while the alternatives are not.  This is particularly
important given greater single- than double-precision throughput on
modern devices, and the consequent temptation to use single precision
with a greater number of particles. Finally, we provide auxiliary
functions useful for implementation, such as for the permutation of
ancestry vectors to enable in-place propagation.
\end{abstract}

\noindent%

\section{Introduction}

The particle filter, and more generally Sequential Monte Carlo (SMC)
methods, constitute a large class of numerical methods routinely used
to perform statistical inference. Particle filters were originally
developed for object tracking and time series analysis using
nonlinear, non-Gaussian state-space models
~\citep{Gordon1993,Doucet2001}.  They have been extended to
accommodate general statistical models
~\citep{Chopin2002,DelMoral2004,DelMoral2006}, with recent
applications including rare event estimation \citep{CerouRareEvent},
graphical models \citep{naesseth2014sequential}, phylogenetic
inference \citep{bouchard2012phylogenetic} and variable selection
\citep{schafer2013sequential}. SMC has shown comparative advantage
over Markov chain Monte Carlo (MCMC) when the target distribution is
multimodal~\citep{chopin2010free,schweizer2012non} or when the
interest lies in the normalizing constant of the target distribution
~\citep{zhou2013towards}.

The general framework of SMC involves introducing a sequence of
distributions $\pi_0,\ldots,\pi_T$, where the interest might be in
each distribution $\pi_t$ or only in the last one $\pi_T$. At step
$t=0$, particles $\mathbf{x}_0^{1:N}\equiv \mathbf{x}_0^{1}, \ldots,
\mathbf{x}_0^{N}$ are drawn independently from $\pi_0$, and each
weight $w_0^i$ in the vector $w_0^{1:N}$ is set to $1/N$. The weighted
particles $(w_0^{1:N}, \mathbf{x}_0^{1:N})$ constitute an empirical
approximation of $\pi_0$.  Then at any step $t\geq 1$ of the
algorithm, the previous particles $(w_{t-1}^{1:N},
\mathbf{x}_{t-1}^{1:N})$ approximating $\pi_{t-1}$ are propagated and
weighted to obtain new particles $(w_t^{1:N}, \mathbf{x}_t^{1:N})$,
which approximate $\pi_t$. A generic way to achieve this
~\citep{DelMoral2004} involves sequences of Markov kernels $K_t$ and
\emph{potential} functions $G_t$ taking values in $(0,+\infty)$, as in
the algorithm in Code \ref{code:smc}.

\begin{code}[ht]
\begin{codebox}
\Procname{$\proc{Sequential-Monte-Carlo}(N \in \{1,2,\ldots\}, T \in \{0,1,\ldots\})$}
\li \Foreach $i \in \{1,\ldots,N\}$
\li   $\mathbf{x}^i_0 \sim \pi_0(\mathbf{x}_0)$ \Comment initialise particle $i$
\li   $w^i_0 \leftarrow 1/N$ \Comment initialise weight $i$
    \End
\li \For $t = 1,\ldots,T$
\li   $\mathbf{a}_t \leftarrow \proc{Ancestors}(\mathbf{w}_{t-1})$ \Comment resample
\li   \Foreach $i \in \{1,\ldots,N\}$
\li     $\mathbf{x}^i_t \sim K_t(\mathbf{x}_t|\mathbf{x}^{a^i_t}_{t-1})$
\Comment propagate particle $i$
\li     $w^i_t \leftarrow G_t(\mathbf{x}_{t}^{i})$ \Comment weight
particle $i$
      \End
    \End
\end{codebox}

\caption{Pseudocode for a generic SMC method.}
\label{code:smc}
\end{code}

Before giving more details on the resampling step, let us describe two
choices for $K_t$ and $G_t$. Consider, for example, a state-space
model where $\mathbf{x}_{0:T}$ is an unobserved Markov chain and
$\mathbf{y}_{1:T}$ are conditionally independent observations of
$\mathbf{x}_{0:T}$ with additional noise, such that the joint
distribution can be written
\begin{equation}
p(\mathbf{x}_{0:t},\mathbf{y}_{1:t}) = p(\mathbf{x}_0)\prod_{t=1}^T
p(\mathbf{y}_t\,|\,\mathbf{x}_t) p(\mathbf{x}_t\,|\,\mathbf{x}_{t-1}).
\end{equation}
For a given data set $\mathbf{y}_{1:T}$, the interest is to draw
samples from the filtering distributions $\pi_t(\mathbf{x}_t) =
p(\mathbf{x}_t\,|\,\mathbf{y}_{1:t})$ for $t = 1,\ldots,T$. When the
probability densities $p(\mathbf{x}_t\,|\,\mathbf{x}_{t-1})$ and
$p(\mathbf{y}_t\,|\,\mathbf{x}_t)$ are linear and Gaussian, the Kalman
filter~\citep{Kalman1960} can be used for this purpose. When they are
nonlinear and non-Gaussian, the particle filter is preferred, as it
provides asymptotically consistent estimates of quantities of interest
as $N \rightarrow \infty$.  The \emph{bootstrap} particle
filter~\citep{Gordon1993} corresponds to the choice
$K_{t}(\mathbf{x}_t | \mathbf{x}_{t-1}) = p(\mathbf{x}_t |
\mathbf{x}_{t-1})$ and $G_t(\mathbf{x}_{t}) = p(\mathbf{y}_t|
\mathbf{x}_t)$.

Another example is that of parameter inference, where the interest is
in the posterior distribution $\pi(\boldsymbol\theta) =
p(\boldsymbol\theta | \mathbf{y}_{1:T})$ of a parameter
$\boldsymbol\theta$ given a data set $\mathbf{y}_{1:T}$. Let
$\pi_0(\boldsymbol{\theta}) = p(\boldsymbol{\theta})$ (the prior
distribution) and introduce, for all $t= 1,\ldots,T$,
$\pi_t(\boldsymbol\theta) \propto p(\boldsymbol\theta)
p(\mathbf{y}_{1:t} | \boldsymbol\theta)$. In this case, a practical
choice~\citep{Chopin2002} is to choose $K_t$ to be an MCMC kernel
leaving $\pi_{t-1}$ invariant, such as a Metropolis--Hastings kernel,
and to define $G_t(\boldsymbol\theta) := p(\mathbf{y}_t |
\mathbf{y}_{1:t-1}, \boldsymbol\theta)$, or, in the case of
independent observations, $G_t(\boldsymbol\theta) := p(\mathbf{y}_t |
\boldsymbol\theta)$.  Since any MCMC kernel can be chosen for $K_t$,
SMC can be seen as a framework to turn an arbitrary MCMC algorithm
into a population-based, and thus parallelisable, algorithm for
parameter inference.

\subsection{Resampling}

In Code \ref{code:smc}, the resampling step is encoded by a randomised
algorithm $\proc{Ancestors}$ that accepts a vector $\mathbf{w}_{t-1}
\in [0,\infty)^N$ of weights, and returns a vector $\mathbf{a}_t \in
  \{1,\ldots,N\}^N$, where each $a^i_t$ is the index of the particle
  at time $t - 1$ which is to be the \emph{ancestor} of the $i$th
  particle at time $t$.  Alternatively, the resampling step may be
  encoded by a randomised algorithm $\proc{Offspring}$ that also
  accepts a vector $\mathbf{w}_{t-1} \in [0,\infty)^N$ of particle
    weights, but instead returns a vector $\mathbf{o}_t \in
    \{0,\ldots,N\}^N$, where each $o^i_t$ is the number of
    \emph{offspring} to be created from the $i$th particle at time $t
    - 1$ for propagation to time $t$.  Ancestry vectors are readily
    converted to offspring vectors and vice-versa (Appendix
    \ref{sec:auxiliary} provides functions to achieve this).

\begin{figure}[t]
\centering
\includegraphics[width=\textwidth]{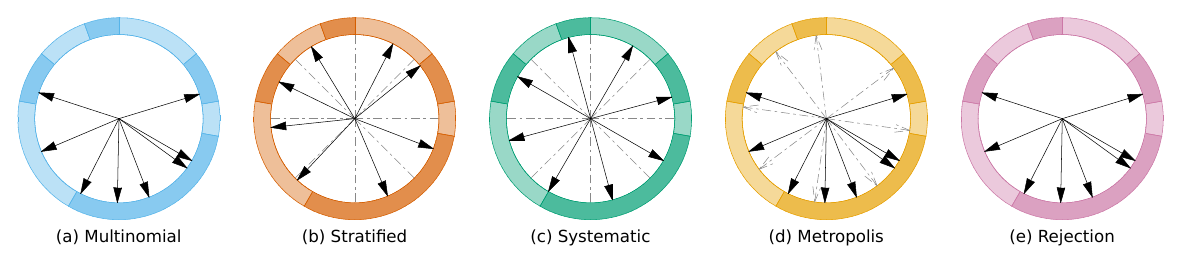}
\caption{Visualisation of the resampling algorithms considered. Arcs
  along the perimeter of the circles represent particles by weight,
  solid arrows indicate selected particles and are positioned
  \textbf{(a)} uniformly randomly in the multinomial resampler,
  \textbf{(b \& c)} by evenly slicing the circle into strata (dashed
  lines) and randomly selecting an offset into each stratum
  (stratified resampler) or using the same offset (systematic
  resampler), \textbf{(d)} by initialising multiple Markov chains
  (dashed arrows) and simulating to convergence in the Metropolis
  resampler, or \textbf{(e)} by rejection sampling. The outputs of
  multinomial, Metropolis and rejection resamplers follow the same
  distribution (represented by identical solid arrows).}
\label{fig:radial}
\end{figure}

There are numerous acceptable algorithms for the resampling
step. Recalling that the output is random, typically it is required
only that, $\forall i \in \{1,\ldots,N\}$:
\begin{equation}\label{eqn:unbiasedness}
\mathbb{E}(o_t^i \mid \mathbf{w}_{t-1}) = \frac{Nw_{t-1}^i}{\sum_{j=1}^N w_{t-1}^j},
\end{equation}
that is, the expected number of offspring of a particle should be
equal to $N$ times its normalised weight. This \emph{unbiasedness
  condition} ensures unbiased estimates of quantities such as the
marginal likelihood $p(\mathbf{y}_{1:T})$, which follows from a simple
extension of Proposition 7.4.1 in \citet{DelMoral2004}. Details of
standard unbiased strategies, including their specific implementation
in this work, are given in Appendix \ref{sec:resampling}. One common
approach is to draw $\mathbf{o}_t$ according to a multinomial
distribution with parameters $N$ and $\mathbf{w}_{t-1}$ (multinomial
resampling). We will consider alternative strategies below, including
some opportunities to relax the unbiasedness condition in exchange for
a significant reduction in execution time. Figure \ref{fig:radial}
visualises both standard and alternative approaches.

\subsection{Parallelisation}

The initialisation, propagation and weighting steps of SMC are readily
parallelised, being independent operations on each particle
$\mathbf{x}_t^i$ and its weight $w_t^i$. Resampling, on the other
hand, is a collective operation across particles and weights, so that
parallelisation is more difficult. Summing the $N$ weights
necessitates synchronization across threads, so that some threads may
wastefully idle while waiting on others. For certain hardware that
does not permit global communication between concurrently running
threads, such as graphics processing units (GPUs), the elimination of
collective operations can yield significant speed up. For this reason
the resampling step has attracted recent attention as a potential
bottleneck in the further scaling of SMC to larger
systems~\citep{Murray2011a,Whiteley2013}.

At the core of most standard resampling schemes, such as the
multinomial, stratified and systematic schemes, is a cumulative sum of
weights, also called a \emph{prefix sum}. A major theme of prior
contributions has been the parallelisation of this prefix
sum~\citep{Maskell2006,Hendeby2010,Chao2010,Gong2012}. This is a
generic problem that is also relevant to other algorithms~\citep[see
  e.g.][]{Harris2007}. Another theme in prior work is the partitioning
of particles into disjoint subsets within which local resampling is
performed~\citep{Chao2010}. This is more useful in a distributed
memory context, as it limits communication between
processes~\citep{Whiteley2013}; it has been considered in this context
before~\citep{Brun2002,Bolic2005}. General comments regarding the
parallelisability of particle filtering algorithms are given in
\citet{Lee2010} and \citet{Murray2013b}.

\subsection{Numerical precision and stability}

As demonstrated later in this work, numerical instabilities can be
apparent in standard resampling schemes when $N$ is large. One million
particles is not an unrealistic number for some contemporary
applications of SMC~\citep[see e.g.][]{Klaas2006,Kitagawa2014}.  In
double-precision, where 15 significant figures (in decimal) are
expected, it is unlikely that $N$ will be sufficiently large for
numerical instability to be a problem in current applications. In
single-precision, however, where 8 significant figures (in decimal)
are expected, numerical instabilities can manifest with this many
particles. This is important because contemporary hardware has
significantly faster single-precision than double-precision
floating-point performance. There is reason to believe that this gap
will remain; consider, for example, that single-precision is twice as
fast on CPU architectures---even reasonably mature
architectures---when SIMD instructions (such as those of SSE and AVX)
are employed. On current GPUs, while the architecture is changing more
rapidly, using single-precision also leads to at least a two-fold
speedup.  On other architectures, such as field-programmable gate
arrays (FPGAs), custom precision is possible, allowing a trade-off
between accuracy and performance ~\citep{Mingas2012}.

\subsection{Contributions}

As identified above, standard resampling algorithms require a
cumulative sum over weights, leading to two problems:
\begin{enumerate}
\item they are less readily parallelised, and
\item they exhibit numerical instability for large numbers of
  particles or large weight variance.
\end{enumerate}
This work contributes two alternative algorithms that eliminate the
cumulative sum over weights in order to remedy both problems. They are
better suited to the breadth of parallelism afforded by modern
hardware, and do not exhibit the numerical instability of standard
schemes. The two alternative schemes are based on Metropolis and
rejection samplers. In comparing these to standard schemes, we
carefully consider the consequences of the choice of resampling scheme
on both the CPU and GPU, considering how numerical stability, bias,
mean squared error and execution time vary across the number of
particles and the variability in their weights. In this respect, the
work constitutes a thorough study of resampling schemes and a useful
guide to the selection and implementation of the most appropriate
algorithm for a given problem.

Section \ref{sec:proposedresampling} describes the Metropolis and
rejection resampling schemes. We also describe a useful permutation of
ancestry vectors in Section \ref{sec:permute} to prevent read and
write conflicts between concurrently running threads.  Empirical
comparisons around bias, mean squared error and execution time are
given in Section \ref{sec:experiments}, with concluding remarks in
Section \ref{sec:conclusion}.

Appendix \ref{sec:conventions} provides our pseudocode conventions for
reference. Appendix \ref{sec:resampling} recalls the standard
algorithms for resampling based on multinomial, stratified and
systematic sampling, highlighting their use of collective
operations. Appendix \ref{app:permute} gives more details on the
permutation algorithm.  Appendix \ref{sec:auxiliary} presents
auxiliary functions for converting between offspring and ancestry
vectors. Appendix \ref{sec:implementation} provides some
implementation notes.

\section{Alternative resampling schemes}\label{sec:proposedresampling}

In this section we describe two approaches, not typically considered
in the literature, that bypass the two issues shared by the standard
schemes. Henceforth, we omit the subscript $t$ from weight, offspring
and ancestry vectors, as the algorithms presented behave identically
at each step of SMC.

\subsection{Metropolis resampling}\label{sec:metropolis}

The first approach resamples via the Metropolis
algorithm~\citep{Metropolis1953} rather than direct sampling, giving a
result close to that of the multinomial resampler. The approach was
briefly studied by the first author in a technical report
\citep{Murray2011a}, but a more complete treatment and improved
analysis is given here. Instead of the collective operation, only the
ratio between pairs of weights is ever computed. Code
\ref{code:metropolis} describes the approach.

\begin{code}[ht]
\begin{codebox}
\Procname{$\proc{Metropolis-Ancestors}(\mathbf{w} \in [0,\infty)^N, B \in \{1,2,\ldots\}) \rightarrow \{1,\ldots,N\}^N$}
\li \Foreach $i \in \{1,\ldots,N\}$
\li     $k \leftarrow i$
\li     \For $n = 1,\ldots,B$
\li         $u \sim \mathcal{U}[0, 1]$
\li         $j \sim \mathcal{U}\{1,\ldots,N\}$ \label{line:proposal}
\li         \If $u \leq w^j/w^k$ \Then \label{line:accept}
\li             $k \leftarrow j$
            \End
        \End
\li     $a^i \leftarrow k$
      \End
    \End
\li \Return $\mathbf{a}$
\end{codebox}

\caption{Pseudocode for Metropolis resampling.}
\label{code:metropolis}
\end{code}

The Metropolis resampler is parameterised by $B$, the number of iterations to
be performed before convergence is assumed and each particle may settle on its
chosen ancestor. We can view the inner $\kw{for}$ loop as iterating a Markov
kernel $P$ with stationary distribution $\pi$ for $B$ steps, where
$$\pi(i) = \frac{w^i}{\sum_{j=1}^N w^j},~i \in \{1,\ldots,N\},$$ is the
categorical distribution associated with the weights.

As $B$ must be finite, the algorithm produces a biased sample. Setting
$B$ is a tradeoff between speed and bias, with smaller $B$ giving
faster execution time but larger bias. This bias may not be much of a
problem for filtering applications, but does violate the assumptions
that lead to unbiased marginal likelihood estimates in a particle MCMC
framework~\citep{Andrieu2010}, so care should be taken.

To provide guidance as to the selection of $B$, we bound the total variation
distance of the $B$-fold iterate $P^B(i,\cdot)$ from $\pi$, where
$$\Vert P^B(i,\cdot) - \pi(\cdot) \Vert_{\rm{TV}} = \max_{A \subseteq
  \{1,\ldots,N\} } |P^B(i,A) - \pi(A)|.$$ Such a bound can be obtained by
noting that $P$ is an independent Metropolis Markov kernel with target $\pi$
and a uniform proposal on $\{1,\ldots,N\}$. By Theorem~2.1 of
\citet{Mengersen1996}
$$\Vert P^B(i,\cdot) - \pi(\cdot) \Vert_{\rm{TV}} \leq (1 - \beta)^B,$$
where
\begin{eqnarray}
\beta &=&
   \min_{i\in\{1,...,N\}}\left(\frac{\frac{1}{N}\sum_{j=1}^{N}w^{j}}{w^{i}}\right) \label{eqn:beta} \geq \frac{1}{N} > 0. \label{eqn:beta-bound}
\end{eqnarray}
Because $\beta > 0$ implies that the associated Markov chain is uniformly
ergodic, and from \citet{Liu1996a}, we know that the spectral gap of $P$ is
exactly $\beta$. To ensure that $\Vert P^B(i,\cdot) - \pi(\cdot)
\Vert_{\rm{TV}} \leq \epsilon$ for a given $\epsilon > 0$ it then suffices to
choose
\begin{equation}\label{eqn:B}
B \geq B^* := \frac{\log\epsilon}{\log\left(1-\beta\right)}.
\end{equation}
This requires a value or lower bound on $\beta$, whose computation we would like to avoid. The bound $1/N$ in (\ref{eqn:beta}) is
too weak, as it leads to setting $B$ roughly as a multiple of $N$ for large $N$. It is sensible instead to choose $\beta$ as some estimate of
\begin{equation}\label{eqn:beta2}
\mathbb{E}(\bar{w})/{w_{\text{max}}},
\end{equation}
where $\bar{w}$ and $w_{\text{max}}$ are respectively the mean of the weights and an upper bound on the weights.
The serial complexity of the Metropolis resampler is $\mathcal{O}(NB)$, but
$B$ may itself be a function of $N$ and the distribution of weights, as in the
analysis above.

\subsection{Rejection resampling}\label{sec:rejection}

When an upper bound on the weights is known \textsl{a priori}, rejection
sampling is possible. Like the Metropolis resampler, the rejection resampler
avoids collective operations and associated numerical instability, but offers
a couple of additional advantages:
\begin{enumerate}
\item it is unbiased,
\item it permits a first deterministic proposal that $a^i = i$, increasing the
  probability of this outcome, and reducing the variance in the ancestry
  vectors produced.
\end{enumerate}
Pseudocode is given in Code \ref{code:rejection}. If line 2 is replaced with
$j\sim \mathcal{U}\{1,\ldots,N\}$ (forgoing the second advantage above),
rejection resampling is an alternative implementation of multinomial
resampling. Its serial complexity is then $\mathcal{O}(N w_{\text{max}} /
\bar{w})$.

\begin{code}[ht]
\begin{codebox}
\Procname{$\proc{Rejection-Ancestors}(\mathbf{w} \in [0,\infty)^N) \rightarrow \{1,\ldots,N\}^N$}
\li \Foreach $i \in \{1,\ldots,N\}$
\li     $j \leftarrow i$
\li     $u \sim \mathcal{U}[0,1]$
\li     \While $u > w^j/w_{\text{max}}$ \label{line:rejection-loop}
\li        $j \sim \mathcal{U}\{1,\ldots,N\}$
\li        $u \sim \mathcal{U}[0,1]$
        \End
\li     $a^i \leftarrow j$
    \End
\li \Return $\mathbf{a}$
\end{codebox}

\caption{Pseudocode for rejection resampling.}
\label{code:rejection}
\end{code}

An issue unique to the rejection resampler is that the computational
effort required to draw each ancestor varies, depending on the number
of rejected proposals before acceptance. This is an example of a
\emph{variable task-length problem}~\citep{Murray2012}, particularly
acute in the GPU context. On GPUs, threads are grouped into warps and
execute the same instructions in parallel. The threads in the same
warp may trip the loop on line \ref{line:rejection-loop} of Code
\ref{code:rejection} a different number of times. All threads in the
warp must complete before any thread can proceed beyond the loop. This
is a particular case of \emph{warp divergence}, which harms
performance. A persistent threads strategy~\citep{Aila2009,Murray2012}
might be used to mitigate the effects of this, although we have not
been successful in finding such an implementation that does not lose
more than it gains through additional overhead in register use and
branching.

If line 2 of Code \ref{code:rejection} is modified so that $j$ is sampled
uniformly on $\{1,\ldots,N\}$ then the number of iterations in the
$\kw{while}$ loop is a geometric random variable with success probability
given by $p := \beta \times \max_i w^i / w_{\text{max}}$, where $\beta$ is the
same as that of (\ref{eqn:beta}), perhaps also chosen as an estimate of
(\ref{eqn:beta2}). The rejection resampler will perform poorly if this
probability is small, which can occur, e.g., when $\max_i w^i \ll
w_{\text{max}}$. One could use the empirical maximum, $w_{\text{max}} = \max
\{w^1,\ldots,w^N\}$, but this would require a collective operation over
weights that would defeat the purpose of the approach and, moreover, $\beta$
could still be small.

Because of the variable task length, the serial complexity of the rejection
resampling algorithm may not be as interesting as its parallel complexity. In
order to determine the expected time to draw $N$ samples, we need to bound the
expectation of the maximum of $N$ independent and identically
distributed geometric random variables, $\mathbb{E}(M_{N}^{*})$. Such a bound
is obtained in \citet{Eisenberg2008}, and we have
\begin{equation}
\frac{1}{\lambda}\sum_{k=1}^{N}\frac{1}{k}\leq\mathbb{E}\left(M_{N}^{*}\right)<1+\frac{1}{\lambda}\sum_{k=1}^{N}\frac{1}{k}.
\end{equation}
Since $H_{N}:=\sum_{k=1}^{N}\frac{1}{k}$ is the $N$th harmonic number, we
additionally have the bounds
\begin{equation}
\frac{1}{2}+\log N<H_{N}<1+\log N
\end{equation}
and so we have that
\begin{equation}
\mathbb{E}\left(M_{N}^{*}\right)<1+\frac{1}{-\log(1-p)}H_{N}<1+\frac{1+\log
  N}{-\log(1-p)}.
\end{equation}
Noting the inequality $-\log(1-p)>p$ for $0<p<1$, we have
$\mathbb{E}\left(M_{N}^{*}\right)<1+\frac{1+\log N}{p}$, and therefore that
the expected parallel complexity of rejection resampling with $N$ processors
is $\mathcal{O}(\frac{\log N}{p})$. This compares to $\mathcal{O}(B) =
\mathcal{O}(\frac{\log\epsilon}{\log(1-p)})<\mathcal{O}(\frac{-\log\epsilon}{p})$
for the Metropolis resampler with $N$ processors and $B$ set to $B^*$ in
(\ref{eqn:B}). This suggests that, if bias in the resampling step is
acceptable and $B$ can be chosen appropriately, Metropolis resampling may be
more appropriate than rejection resampling.

Performance can be tuned if one is willing to concede a weighted outcome from
the resampling step, rather than the usual unweighted outcome. This is the
approach taken with the \emph{partial rejection control}
heuristic~\citep{Liu1998a}. To do this, choose some $v_{\text{max}} <
w_{\text{max}}$, then form a categorical distribution using the weights
$v^1,\ldots,v^N$, where $v^i = \min (w^i, v_{\text{max}})$. Clearly
$v_{\text{max}}$ forms an upper bound on these new weights. One could sample
from this using Code \ref{code:rejection}, with $v$ in place of $w$, and then
importance weight each particle $i$ with $w^i \leftarrow
w^{a^i}/v^{a^i}$. Note that each weight is 1 except where $w^{a^i} >
w_{\text{max}}$. The procedure may also be used when no hard upper bound on
weights exists ($w_{\text{max}}$), but where some reasonable substitute can be
made ($v_{\text{max}}$).

\subsection{Ancestor permutation for in-place propagation}\label{sec:permute}

A desirable feature of a resampling scheme is for it to allow in-place
propagation of the particle system. This is useful for
memory-intensive applications of SMC in general
\citep{bouchard2012phylogenetic}, but particularly for GPU
implementations, since the memory available to GPU devices is
typically much smaller than main memory. Instead of having an input
buffer holding particles at time $t - 1$, and a separate output buffer
into which to write the propagated particles at time $t$, a single
buffer is used with the time $t$ particles replacing the time $t - 1$
particles. This in-place operation is more memory efficient by a
factor of two (for a fixed number of particles), but requires
guarantees that the reads and writes on the single buffer can be
executed concurrently without conflicts. That is, each particle is
either read from or written to, but not both.

To work in-place and prevent read and write conflicts, it is
sufficient that the ancestry vector, $\mathbf{a}_t$, satisfies
$\forall i \in \{1,\ldots,N\}$:
\begin{equation}\label{eqn:predicate}
o_t^i > 0 \implies a_t^i = i.
\end{equation}
With this, it is possible to insert a copy step immediately before
each propagation step, setting $\mathbf{x}_{t-1}^i \leftarrow
\mathbf{x}_{t-1}^{a_t^i}$ for all $i \in \{1,\ldots,N\}$ where $a_t^i \neq
i$. Each particle can then be propagated in-place by reading from and
writing to the same buffer. 
The ancestry vector produced by resampling schemes will not typically
satisfy \eqref{eqn:predicate}, but a permutation of it will.  In
Appendix \ref{app:permute} we describe an algorithm to re-order the
ancestry vector to achieve this, in both serial and parallel settings.


\section{Results and discussion}\label{sec:experiments}

The resampling algorithms are assessed empirically for bias, mean
squared error and execution time.  Single precision is used for all
experiments, in order to highlight some of the numerical issues
arising when standard resampling schemes are used with large numbers
of particles. While the use of double precision eliminates the
numerical artifacts in the results, the ranking of algorithms by
execution time is unaffected.

Experiments are conducted on two devices. The first is an eight-core
Intel Xeon E5-2650 CPU, compiling with the Intel C++ Compiler version
12.1.3, using OpenMP to parallelise over eight threads. The second
device is an NVIDIA K20 GPU hosted by the same CPU, compiling with
CUDA 5.0 and the same version of the Intel compiler. All compiler
optimisations are applied. In particular, we use the \texttt{-arch
  sm\_35} option to the CUDA compiler to target the specific
architecture of the NVIDIA K20.

\subsection{Framework}

Resampling algorithms are often assessed using the mean squared error
(MSE, see e.g. \citet{Kitagawa1996}),  computed from the offspring
vector $\mathbf{o}$ and weight vector $\mathbf{w}$. The squared error (SE) of
a particular offspring vector $\mathbf{o}_k$ is:
\begin{equation}
\mathrm{SE}(\mathbf{o}_k) = \sum_{i=1}^N\left(o_k^i - \frac{Nw^i}{\proc{Sum}(\mathbf{w})}\right)^2.
\end{equation}
For some set of $K$ offspring vectors, $\{\mathbf{o}_1,\ldots,\mathbf{o}_K\}$,
the mean squared error is simply the sample mean of these squared errors:
$\mathrm{MSE}(\mathbf{o}) = \frac{1}{K}\sum_{k=1}^K\mathrm{SE}(\mathbf{o}_k)$.
The MSE can be written as separate bias and variance components:
\begin{equation}\label{eqn:mse}
\mathrm{MSE}(\mathbf{o}) = \mathrm{tr}\left(\mathrm{Var}(\mathbf{o})\right) +
\|\mathrm{Bias}(\mathbf{o})\|^2,
\end{equation}
noting:
\begin{align*}
\mathrm{tr}\left(\mathrm{Var}(\mathbf{o})\right) = \sum_{i=1}^N \mathrm{Var}(o^i) \quad&\text{and}\quad
\|\mathrm{Bias}(\mathbf{o})\|^2 = \sum_{i=1}^N
\left(\hat{o}^i - \frac{Nw^i}{\proc{Sum}(\mathbf{w})}\right)^2,
\end{align*}
where $\hat{o}^i$ denotes the sample mean of the $i$th component of
$\mathbf{o}$ across the set of $K$ offspring vectors, and
$\mathrm{Var}(o^i)$ the sample variance of the same. This permits
separate evaluation of the bias of each resampling algorithm,
recalling the unbiasedness condition (\ref{eqn:unbiasedness}). This is
particularly important for the Metropolis resampler, which is biased
for any finite number of steps, $B$. The other algorithms have zero
bias in theory, but may exhibit bias in practice due to numerical
issues. Algorithms are assessed below using the contribution of the
squared bias to the
MSE $$\|\mathrm{Bias}(\mathbf{o})\|^2/\mathrm{MSE}(\mathbf{o}),$$ as
well as the MSE normalised by the number of particles,
$\mathrm{MSE}(\mathbf{o})/N$.

Weight sets are simulated to assess the speed and accuracy of each resampling
algorithm. For a number of particles $N$ and observation $y$, a weight set is
generated by sampling $x^i \sim \mathcal{N}(0,1)$, for $i = 1,\ldots,N$, and
setting
\begin{equation}
w^i = \frac{1}{\sqrt{2\pi}}\exp\left(-\frac{1}{2}(x^i - y)^2\right).
\end{equation}
The construction is analogous to having a prior distribution of $x \sim
\mathcal{N}(0,1)$ and likelihood function of $y \sim \mathcal{N}(x,1)$. As $y$
increases, the relative variance in weights does too. For this set up, the
maximum weight is $w_{\text{max}} = 1/\sqrt{2\pi}$,
and the expected weight
\begin{equation}
\mathbb{E}(w) = \mathcal{N}(y;0,\sigma^2=2) = \frac{1}{2\sqrt{\pi}}\exp\left(-\frac{1}{4}y^2\right).
\end{equation}
The variance of the weights is
\begin{equation}
\mathbb{V}(w) = \frac{1}{\pi\sqrt{12}}\exp\left(-\frac{1}{3}y^2\right) - [\mathbb{E}(w)]^2,
\end{equation}
and thus their \emph{relative} variance  $\mathbb{V}(w)/\mathbb{E}(w)$ is increasing with $y$.

These are used to set the number of steps for the Metropolis resampler
according to the analysis in Section \ref{sec:metropolis}. Using
$\epsilon = 1/100$ and $\beta = \bar{w}/w_{\text{max}}$, we set $B =
B^*$ as defined in (\ref{eqn:B}). The maximum weight $w_{\text{max}}$
is also used for the rejection resampler.
This procedure is used to generate 16 different weight vectors for each
combination of $N = 2^4,2^5,\ldots,2^{22}$ and $y =
0,\frac{1}{2},1,1\frac{1}{2},\ldots,4$. For each of these 16 weight vectors,
each resampling algorithm is used to draw 256 offspring vectors. Results
reported below are averages over the 16 weight vectors.

\subsection{Bias results}

\begin{figure}[t]
\centering
\includegraphics[width=\linewidth]{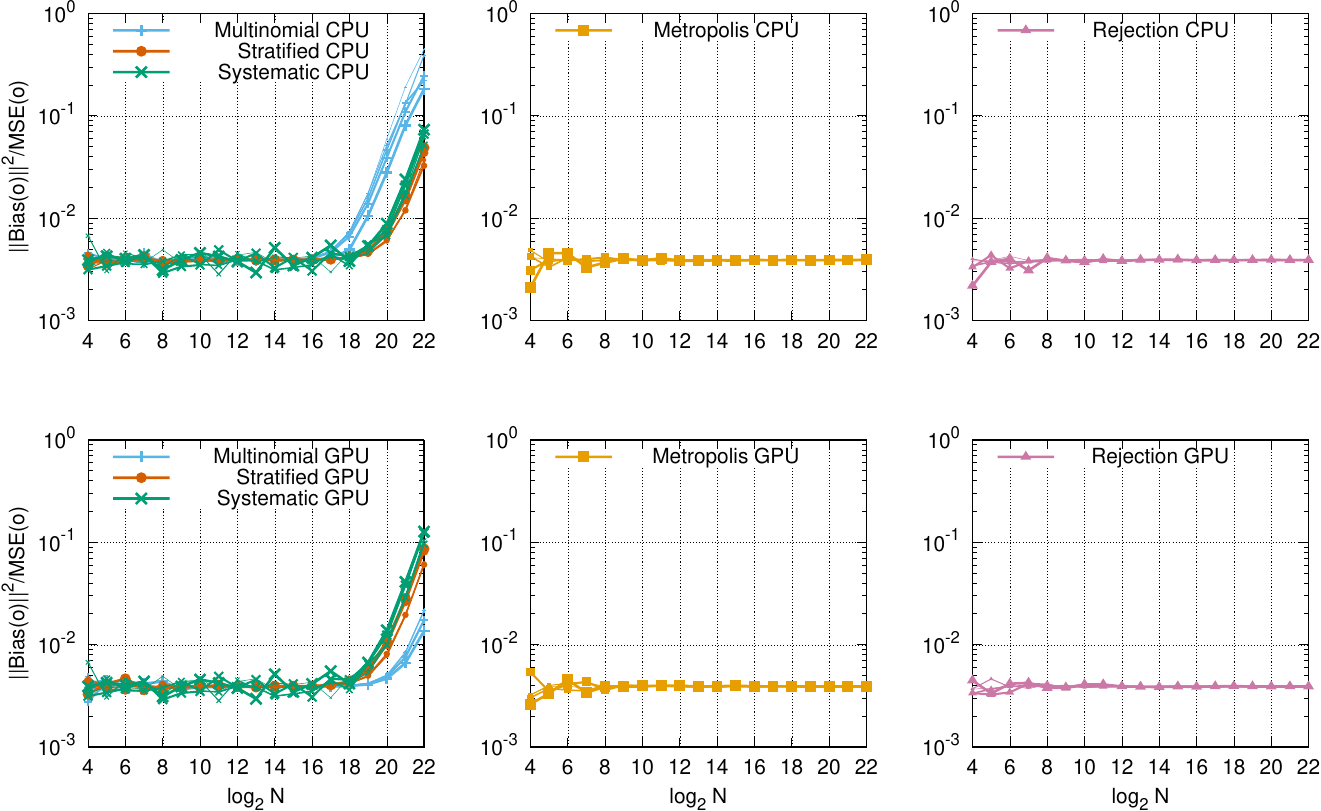}
\caption{Contribution of the bias to the MSE for the various
  resampling algorithms executed in single precision floating point
  arithmetic on \textbf{(top row)} CPU and \textbf{(bottom row)}
  GPU. For each algorithm, multiple lines of increasing thickness
  indicate results for $y = 0,1,2,3,4$. The increasing bias for the
  multinomial, stratified and systematic resamplers after about $N >
  2^{18}$ is due to numerical instability in the large summations that
  these algorithms require. The bias remains apparent when pre-sorting
  weights, but not when using double precision arithmetic. The
  instability is noticeably worse for the different multinomial
  algorithm used on CPU (Code \ref{code:multinomial-serial} in
  Appendix \ref{sec:multinomial}) than that on GPU (Code
  \ref{code:multinomial} in Appendix \ref{sec:multinomial}).}
\label{fig:bias}
\end{figure}

Figure \ref{fig:bias} plots the contribution of the empirical bias to
the MSE for all algorithms. For the multinomial, stratified and
systematic resamplers, this appears satisfactory until $N \geq
2^{19}$, after which the bias contribution increases rapidly. This is
due to the numerical instability of the cumulative sum required by
these algorithms.  The instability is noticeably worse for the
different multinomial algorithm used on the CPU (Code
\ref{code:multinomial-serial} in Appendix \ref{sec:multinomial}) than
that on the GPU (Code \ref{code:multinomial} in Appendix
\ref{sec:multinomial}); this is explained by the cumulative sum in the
former being linear over a vector rather than recursive over a binary
tree. The Metropolis and rejection algorithms do not share this
instability, and otherwise empirically match the bias contribution of
the other methods. This is to be expected for the rejection
algorithm. For the Metropolis algorithm it suggests that the procedure
for setting $B$ in Section \ref{sec:metropolis} is appropriate. It
also suggests that while the Metropolis algorithm is theoretically
biased, this bias is negligible in practice compared to numerical
errors.

We observe, but do not show, that the pre-sorting of weights does not fix the
numerical instability of the multinomial, stratified and systematic resamplers
for large numbers of particles, but that the use of double precision does. 
The number of particles that would be required to reproduce the same instability
in double precision far surpasses, by orders of magnitude, that which would be
realistic to use in SMC at present.

\begin{figure}[t]
\centering
\includegraphics[width=\linewidth]{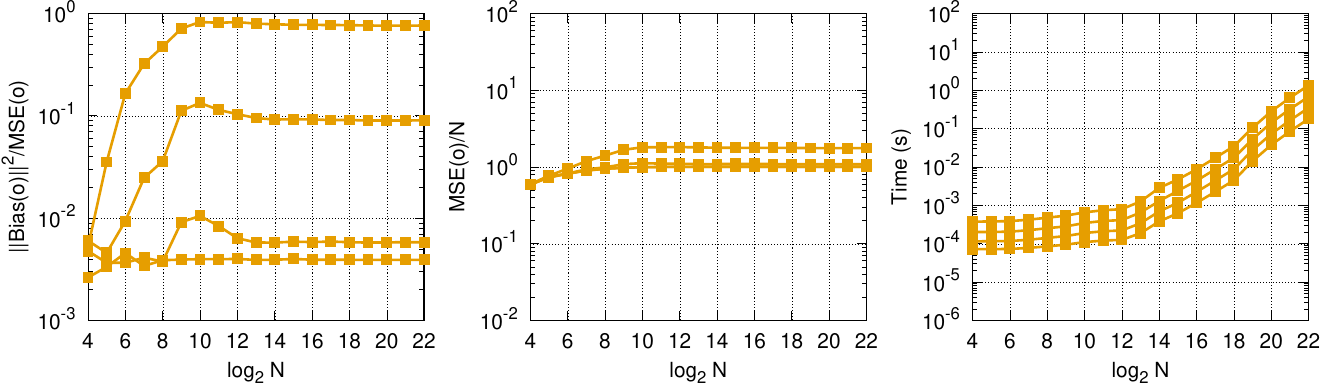}
\caption{The effect of manipulating the number of steps, $B$, in the
  Metropolis resampler: \textbf{(left)} bias, \textbf{(middle)} MSE and
  \textbf{(right)} execution time. In all cases $y = 4$, with $B$ set to
  $B^*/C$ for each $C \in \{1,2,4,8\}$. As $C$ increases, bias increases, MSE
  increases slightly (on account of the bias increase), and execution time
  decreases roughly linearly.}
\label{fig:metropolis}
\end{figure}

The selection of $B$ for the Metropolis resampler appears sufficient, but we
may question whether it is too conservative. To test this empirically we
compare runs of the Metropolis resampler with reduced number of steps, setting
$B = B^*/C$ for each $C \in \{1,2,4,8\}$. The contribution of the empirical
bias to the MSE is given in the leftmost plot of Figure
\ref{fig:metropolis}. As it matches that of the multinomial resampler for $C =
1$---which cannot be improved upon---and noticeably increases for $C \geq 2$,
this suggests that the setting $B = B^*$ is indeed appropriate.

\subsection{MSE results}

\begin{figure}[t]
\centering
\includegraphics[width=\linewidth]{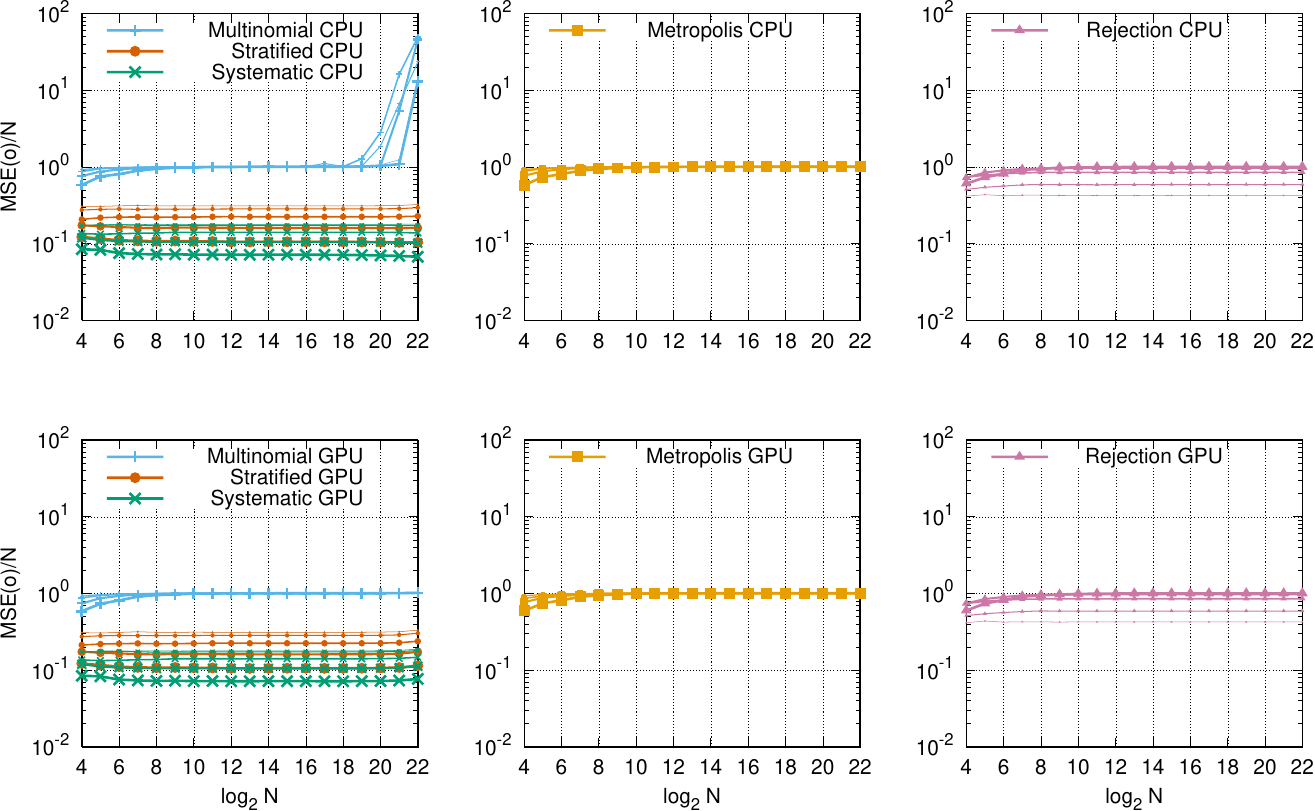}
\caption{MSE of the various resampling algorithms executed in single
  precision floating point arithmetic on \textbf{(top row)} CPU and
  \textbf{(bottom row)} GPU. For each algorithm, multiple lines of increasing
  thickness indicate results for $y = 0,1,2,3,4$.}
\label{fig:var}
\end{figure}

Figure \ref{fig:var} indicates that MSE differs between
methods. Little in this figure is surprising, however: it is well
known that the stratified resampler reduces variance over the
multinomial resampler, and that the systematic resampler can, but does
not necessarily, reduce it again~\citep{Douc2005}. Numerical
instabilities in the CPU implementation of the multinomial resampler
(Code \ref{code:multinomial-serial} in Appendix \ref{sec:multinomial})
appear to increase the MSE in its outcomes for large $N$. Also of
interest is that as $y$ increases, the probability of accepting the
initial proposal of the rejection resampler declines, so that its MSE
degrades away from that of the systematic and stratified resamplers,
towards that of the multinomial and Metropolis resamplers.

\subsection{Execution time results}

\begin{figure}[t]
\centering
\includegraphics[width=\linewidth]{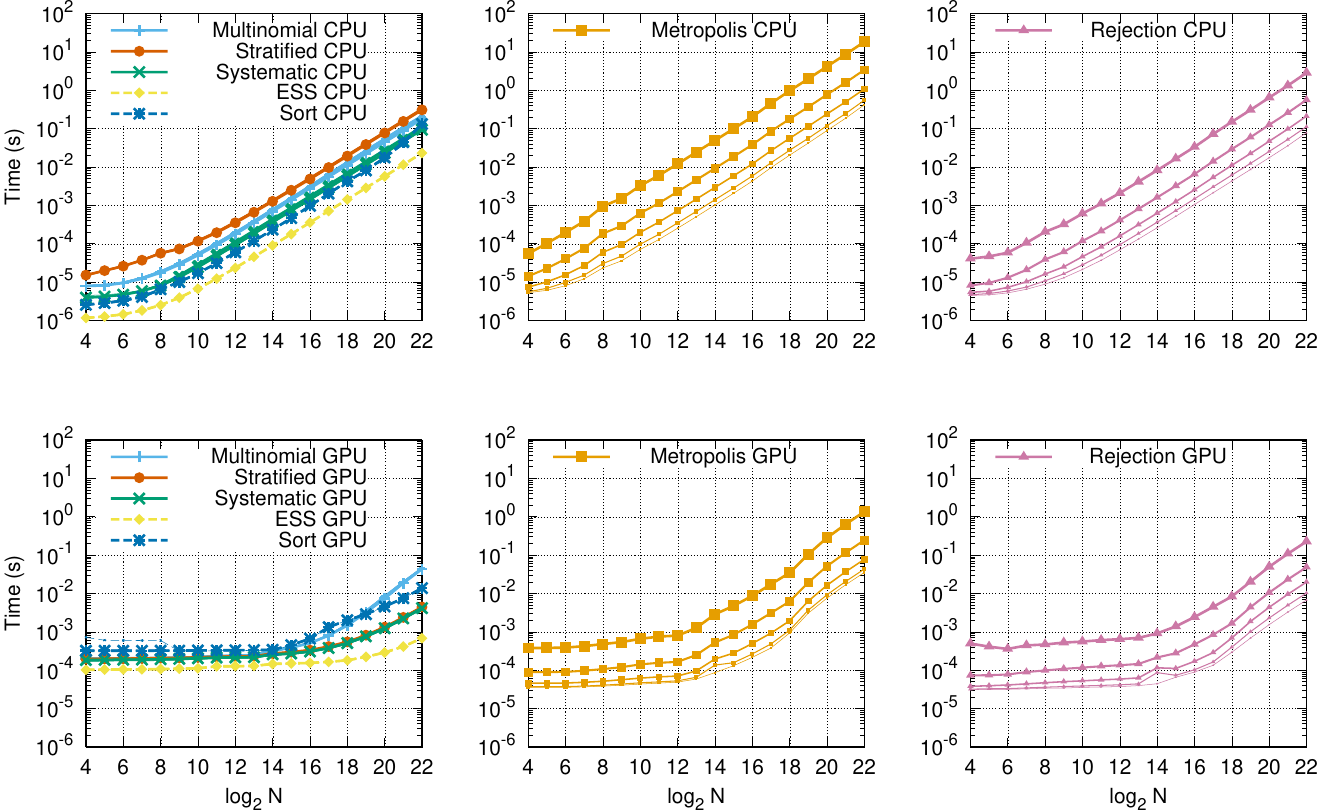}
\caption{Execution times of the various resampling algorithms executed in
  single precision floating point arithmetic on \textbf{(top row)} CPU and
  \textbf{(bottom row)} GPU. For context, the execution times of procedures to
  compute ESS and to pre-sort weights are also included. For each algorithm,
  multiple lines of increasing thickness indicate results for $y =
  0,1,2,3,4$. }
\label{fig:times}
\end{figure}

Figure \ref{fig:times} shows the execution times for all algorithms,
as well as, for context, the execution times of procedures for sorting
a weight vector and computing its \emph{effective sample size}
(ESS)~\citep{Liu1995}. The ESS is given by
$\proc{Sum}(\mathbf{w})^2/\mathbf{w}^T\mathbf{w}$.

Execution times are taken until the delivery of an ancestry vector
satisfying (\ref{eqn:predicate}), and so include any of the auxiliary
functions in Appendices \ref{sec:auxiliary} and \ref{app:permute}
necessary to achieve this. Note that---as we would expect---the
multinomial, stratified and systematic resamplers are not sensitive to
$y$ (or equivalently to the variance in weights) with respect to
execution time, while the Metropolis and rejection resamplers are.

\begin{figure}[t]
\centering
\includegraphics[width=\textwidth]{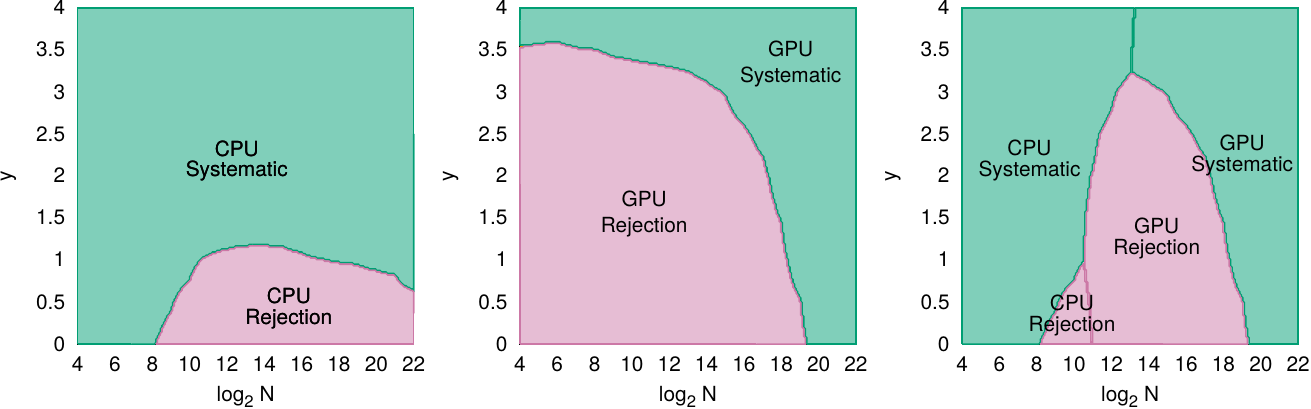}
\caption{Decision matrices for algorithmic choice based on fastest execution
  time. Each matrix shows the algorithm to choose for each combination of
  number of particles, $N$, and weight variability, which increases with
  $y$. Decisions are between \textbf{(left)} all algorithms on CPU only
  \textbf{(centre)} all algorithms on GPU only and \textbf{(right)} all
  algorithms on both CPU and GPU. The inclusion of copying weights and
  ancestry vectors between CPU and GPU does not change the decision boundaries
  significantly in the rightmost plot.}
\label{fig:decisions}
\end{figure}

In a Bayesian decision theoretic setting, we can adopt execution time as a
loss function, and choose, for any combination of $N$ and $y$, the algorithm
that minimises the expectation of this loss function. A more sophisticated
loss function might include the bias and variance as well, but the relative
weighting of the individual components is a subjective decision for the
problem at hand, so we do not attempt to do this. Using execution time alone
as a loss function, Figure \ref{fig:decisions} plots the resulting decision
matrices across all combinations of $N$ and $y$ for which empirical results
were recorded. From these matrices and Figure \ref{fig:times}, we can conclude:
\begin{enumerate}
\item that the GPU should generally not be considered for resampling with
  fewer than $2^{10}$ particles,
\item that the systematic resampler is a good candidate overall, but
\item that there is a significant region of the space, especially at lower
  weight variances, for which the rejection or Metropolis resamplers are
  faster.
\end{enumerate}

We observe, but do not show, that the decision boundaries in Figure
\ref{fig:decisions} are not significantly affected by including the time taken
to copy between GPU device and main memory. This means that the choice between
the CPU or GPU device for resampling is largely independent of the choice of
device for the propagation and weighting of particles. For example, use of the
GPU for propagating and weighting particles does not then greatly favour the
GPU for resampling: the penalty to copy the weight vector to main memory,
resample using the CPU, and then copy the resulting ancestry vector back to
device memory, is not significant.

The choice of $B$ for the Metropolis algorithm permits a trade off
between bias and execution time. This may be particularly useful in
applications with hard execution time constraints, such as real-time
object tracking. Recall that execution time is linear in
$B$. Execution time results for various choices of $B$ are given in
the rightmost plot of Figure \ref{fig:metropolis}, with associated
biases in the leftmost plot. Recall that the rejection resampler is
also somewhat configurable by using an approximate maximum weight. To
do this, one must be willing to accept a still-weighted output from
the resampling step, and the cumulative implications of this within
SMC are problem-specific and not overly clear. We leave this for
future work.

A further consideration is that the execution time of both the
Metropolis and rejection resamplers depends on the PRNG used. This
dependence is by a constant factor, but can be substantial. Here,
robust PRNGs for Monte Carlo work have been used (see Appendix
\ref{sec:implementation}), but conceivably cheaper, if less robust,
PRNGs might be considered. This represents another trade-off between
execution time and bias.

\section{Conclusion}\label{sec:conclusion}

This work has presented two alternative resampling schemes for
SMC that eliminate collective operations over
weights. Consequently, they are more readily parallelised and are more
numerically stable than standard resampling schemes.

The appropriate choice of resampling algorithm
depends on a number of problem-specific factors, including:
\begin{enumerate}
\item the number of particles required and the typical variability in their
  associated weights,
\item whether a maximum weight exists, or can be approximated sufficiently
  accurately, to configure and use the Metropolis and rejection algorithms,
\item the tolerable level of bias in resampling outcomes, and
\item the available numerical precision.
\end{enumerate}
The Metropolis and rejection algorithms:
\begin{enumerate}
\item are demonstrably, in certain circumstances, superior in terms of
  execution time and numerical stability (see e.g. Figures
  \ref{fig:decisions} and \ref{fig:var}, respectively), and
\item allow a practitioner to implement SMC in single precision for
  large numbers of particles without the numerical instabilities of
  existing resampling schemes.
\end{enumerate}

With respect to numerical stability, especially in single precision,
great care should be taken when using the standard multinomial,
stratified or systematic resamplers with upwards of hundreds of
thousands of particles, even though these schemes are unbiased in
theory. This is due to numerical instability in the cumulative sum
operation that these algorithms require. The alternative Metropolis
and rejection resamplers have better numerical properties, as they
compute only ratios of weights. This is important in light of the
temptation to use single-precision, or even custom-precision floating
point to improve execution times with modern computer architectures. 


\section*{Supplementary materials}
\begin{description}
\item[LibBi package for numerical results:] LibBi package
  \texttt{Resampling} providing scripts to reproduce the numerical
  results of this article using all algorithms as implemented in the
  LibBi software, available at
  \url{www.libbi.org}.
\end{description}

\section*{Acknowledgements}

The third author acknowledges EPSRC for funding this research
through grant EP/K009362/1.

\bibliographystyle{chicago}
{\small
\bibliography{gpupf}
}

\appendix

\section{Pseudocode conventions}\label{sec:conventions}

The algorithms presented in this work are described using pseudocode
with a number of conventions. We distinguish between the $\Foreach$
and $\For$ constructs. The former is used where the body of the loop
is to be executed for each element of a set, with the order
unimportant. The latter is used where the body of the loop is to be
executed for each element of a sequence, where the order must be
preserved. The intended implication is that $\Foreach$ loops may be
parallelised, while $\For$ loops cannot be. The $\Atomic$ keyword is
used to indicate that a line must be executed as if it constitutes one
instruction (i.e. an atomic operation) in order to avoid read and
write conflicts between concurrently running threads.

A number of primitive operations such as searches, transformations,
reductions, sorts and prefix sums are used throughout
pseudocode. These are specified in Code \ref{code:primitives}.
Such operations will be familiar to users of, for example, the C++ standard
template library (STL) or Thrust library~\citep{Thrust}, and their
implementation on GPUs has been well-studied~\citep[see
  e.g.][]{Harris2007,Satish2009}. The advantage of describing
algorithms in this way is that we can specify intent without
prescribing implementation; the efficient implementation of these
primitives in both serial and parallel contexts is well understood,
and a single pseudocode description that uses primitives will often
suffice for both serial and parallel contexts. 

\begin{code}[ht]
\begin{codebox}
\Procname{$\proc{Inclusive-Prefix-Sum}(\mathbf{w} \in [0,\infty)^N) \rightarrow [0,\infty)^N$}
\li $W^i \leftarrow \sum_{j=1}^i w^j$
\li \Return $\mathbf{W}$
\end{codebox}

\begin{codebox}
\Procname{$\proc{Exclusive-Prefix-Sum}(\mathbf{w} \in [0,\infty)^N) \rightarrow [0,\infty)^N$}
\li $W^i \leftarrow \begin{cases} 0 & i = 1\\ \sum_{j=1}^{i-1} w^j & i > 1\end{cases}$
\li \Return $\mathbf{W}$
\end{codebox}

\begin{codebox}
\Procname{$\proc{Adjacent-Difference}(\mathbf{W} \in [0,\infty)^N) \rightarrow [0,\infty)^N$}
\li $w^i \leftarrow \begin{cases}W^i & i = 1\\W^i - W^{i-1} & i > 1\end{cases}$
\li \Return $\mathbf{w}$
\end{codebox}

\begin{codebox}
\Procname{$\proc{Sum}(\mathbf{w} \in [0,\infty)^N) \rightarrow [0,\infty)$}
\li \Return $\sum_{i=1}^N w^i$
\end{codebox}

\begin{codebox}
\Procname{$\proc{Lower-Bound}(\mathbf{W} \in [0,\infty)^N, u \in [0,\infty))
    \rightarrow \{1,\ldots,N\}$}
\li \Requires
\li   $W$ is sorted in ascending order
    \End
\li \Return\Indentmore
\li   the lowest $j$ such that $u$ may be inserted into
\zi   position $j$ of $W$ and maintain its sorting.
    \End
\end{codebox}

\caption{Pseudocode for various primitive functions.}
\label{code:primitives}
\end{code}

\section{Standard resampling schemes\label{sec:resampling}}

\subsection{Multinomial resampling}\label{sec:multinomial}

Multinomial resampling proceeds by drawing each $a^i$ independently from the
categorical distribution over $\mathcal{C} = \{1,\ldots,N\}$, where $P(a^i =
j) = w^j/\proc{Sum}(\mathbf{w})$. Pseudocode is given in Code
\ref{code:multinomial}. The algorithm is dominated by the $N$ calls of
\proc{Lower-Bound}, which if implemented with a binary search, will give a
serial complexity of $\mathcal{O}(N \log_2 N)$ overall.

\begin{code}[ht]
\begin{codebox}
\Procname{$\proc{Multinomial-Ancestors}(\mathbf{w} \in [0,\infty)^N)
  \rightarrow \{1,\ldots,N\}^N$}
\li $\mathbf{W} \leftarrow \proc{Inclusive-prefix-sum}(\mathbf{w})$ \label{line:multinomial-prefix-sum}
\li \Foreach $i \in \{1,\ldots,N\}$
\li     $u^i \sim \mathcal{U}[0, W^N)$
\li     $a^i \leftarrow \proc{Lower-bound}(\mathbf{W}, u^i)$
    \End
\li \Return $\mathbf{a}$
\end{codebox}

\caption{Pseudocode for parallel multinomial resampling.}
\label{code:multinomial}
\end{code}

The \proc{Inclusive-Prefix-Sum} operation on line
\ref{line:multinomial-prefix-sum} of Code \ref{code:multinomial} is
not numerically stable, as large values may be added to relatively
insignificant ones during the procedure (an issue intrinsic to any
large summation). With large $N$, assigning the weights to the leaves
of a binary tree and summing with a depth-first recursion over this
will help. With large variance in weights, pre-sorting may also
help. While log-weights are often used in the implementation of SMC,
these need to be exponentiated (perhaps after rescaling) for the
\proc{Inclusive-Prefix-Sum} operation, so this does not alleviate the
issue.

Serially, the same approach may be used, although a single-pass
approach of complexity $\mathcal{O}(N)$ is enabled by generating
sorted uniform random variates~\citep{Bentley1979}. Code
\ref{code:multinomial-serial} details this approach. A drawback is the
use of relatively expensive logarithm functions. There is scope for a
small degree of parallelism in this new algorithm by dividing $N$
among a handful of threads. Each thread must still step through all
$N$ weights, however, so that the complexity is not improved with
parallelism. We find it faster than Code \ref{code:multinomial} when
on CPU, but slower when on GPU.

\begin{code}[ht]
\begin{codebox}
\Procname{$\proc{Multinomial-Ancestors}(\mathbf{w} \in [0,\infty)^N)
  \rightarrow \{1,\ldots,N\}^N$}
\li $\mathbf{W} \leftarrow \proc{Exclusive-prefix-sum}(\mathbf{w})$
\li $W \leftarrow W^N + w^N$ \Comment sum of weights
\zi
\li $\id{lnMax} \leftarrow 0$
\li $j \leftarrow N$
\li \For $i = N,\ldots,1$
\li   $u \sim \mathcal{U}[0,1)$
\li   $\id{lnMax} \leftarrow \id{lnMax} + \ln(u)/i$
\li   $u \leftarrow W\exp(\id{lnMax})$
\li   \While $u < W^j$
\li     $j \leftarrow j - 1$
      \End
\li   $a^i \leftarrow j$
    \End
\li \Return $\mathbf{a}$
\end{codebox}

\caption{Pseudocode for serial, single-pass multinomial resampling.}
\label{code:multinomial-serial}
\end{code}

\subsection{Stratified resampling}\label{sec:stratified}

The variance in outcomes produced by the multinomial resampler may be
reduced~\citep{Douc2005} by stratifying the cumulative probability
function of the same categorical distribution, and randomly drawing
one particle from each stratum. This stratified
resampler~\citep{Kitagawa1996} most naturally delivers not the
ancestry vector $\mathbf{a}$ or offspring vector $\mathbf{o}$, but the
\emph{cumulative offspring} vector, which we denote $\mathbf{O}$, and
define as $\mathbf{O} =
\proc{Inclusive-Prefix-Sum}(\mathbf{o})$. Pseudocode is given in Code
\ref{code:stratified}. The algorithm is of serial complexity
$\mathcal{O}(N)$.

\begin{code}[ht]
\begin{codebox}
\Procname{$\proc{Stratified-Cumulative-Offspring}(\mathbf{w} \in [0,\infty)^N)
  \rightarrow \{0,\ldots,N\}^N$}
\li $\mathbf{u} \sim \text{i.i.d.} \,\, \mathcal{U}[0, 1)$
\li $\mathbf{W} \leftarrow \proc{Inclusive-Prefix-Sum}(\mathbf{w})$ \label{line:stratified-prefix-sum}
\li \Foreach $i \in \{1,\ldots,N\}$
\li     $r^i \leftarrow \frac{NW^i}{W^N}$
\li     $k^i \leftarrow \min\left(N, \left\lfloor r^i \right\rfloor + 1\right)$
\li     $O^i \leftarrow \min\left(N, \left\lfloor r^i + u^{k^i}\right\rfloor\right)$\label{line:stratified-numerical-problem}
    \End
\li \Return $\mathbf{O}$
\end{codebox}

\caption{Pseudocode for stratified resampling.}
\label{code:stratified}
\end{code}

As for multinomial resampling, the \proc{Inclusive-Prefix-Sum} operation on
line \ref{line:stratified-prefix-sum} of Code \ref{code:stratified} is not
numerically stable. The same strategies to ameliorate the problem apply.
Line \ref{line:stratified-numerical-problem} of Code \ref{code:stratified} is
more problematic. Consider that there may be a $j$ such that, for $i \geq j$,
$u^{k^i}$ is not significant against $r^i$ under the floating-point model, so
that the result of $r^i + u^{k^i}$ is just $r^i$. For such $i$, no random
sample is being made within the strata. Furthermore, rounding up on the same
line might easily deliver $O^N = N + 1$, not $O^N = N$ as required, if not for
the quick-fix use of $\min$.
Given that single precision has about seven significant figures in
decimal, consider that, with $N$ around one million, almost certainly
no $u^{k^i} \in [0,1)$ is significant against $r^i$ at high $i$. 
Note that while pre-sorting weights and summing over a binary tree can help
with the numerical stability of the \proc{Inclusive-Prefix-Sum}
operation, it does not help with this latter issue.

\subsection{Systematic resampling}\label{sec:systematic}

The variance in outcomes of the stratified resampler may often, but not
always~\citep{Douc2005}, be further reduced by using the same random offset
within each stratum. This is the \emph{systematic resampler} (equivalent to
the \emph{deterministic} method described in the appendix of
\citet{Kitagawa1996}). Pseudocode is given in Code \ref{code:systematic}, which
is a simple modification to Code \ref{code:stratified}. The same complexity
and numerical caveats apply to the systematic resampler as for the stratified
resampler.

\begin{code}[ht]
\begin{codebox}
\Procname{$\proc{Systematic-Cumulative-Offspring}(\mathbf{w} \in [0,\infty)^N)
  \rightarrow \{0,\ldots,N\}^N$}
\li $u \sim \mathcal{U}[0, 1)$
\li $\mathbf{W} \leftarrow \proc{Inclusive-Prefix-Sum}(\mathbf{w})$
\li \Foreach $i \in \{1,\ldots,N\}$
\li     $\id{r^i} \leftarrow \frac{NW^i}{W^N}$
\li     $O^i \leftarrow \min\left(N, \left\lfloor\id{r^i} + u\right\rfloor\right)$
    \End
\li \Return $\mathbf{O}$
\end{codebox}

\caption{Pseudocode for systematic resampling.}
\label{code:systematic}
\end{code}

The resampling algorithms presented here do not constitute an exhaustive list
of those in use, for instance \emph{residual
  resampling} has been omitted ~\citep{Liu1998}. However they are reasonably representative, and can form the building
blocks of more elaborate schemes.

\section{Ancestor permutation for in-place propagation}\label{app:permute}

An ancestry vector may be permuted to satisfy \eqref{eqn:predicate} in the main
article. A serial algorithm to achieve this is straightforward and
given in Code \ref{code:ancestor-permute-sequential}. This
$\mathcal{O}(N)$ algorithm makes a single pass through the ancestry
vector with pair-wise swaps to satisfy the condition.

\begin{code}[ht]
\begin{codebox}
\Procname{$\proc{Permute}(\mathbf{a} \in \{1,\ldots,N\}^N)$}
\li \For $i = 1,\ldots,N$
\li   \If $a^i \neq i$ and $a^{a^i} \neq a^i$
\li     \Then
          $\func{swap}(a^i, a^{a^i})$
\li       $i \leftarrow i - 1$ \Comment repeat for new value
        \End
    \End
\li \Ensures
\li    $\forall i (i \in \{1,\ldots,N\}: o^i > 0 \implies a^i = i)$
    \End
\end{codebox}

\caption{Serial algorithm for permuting an ancestry to ensure
    condition \eqref{eqn:predicate} in the main article.}
\label{code:ancestor-permute-sequential}
\end{code}

The simple algorithm is complicated in a parallel context as the
pair-wise swaps are not readily serialised without heavy-weight mutual
exclusion. In parallel we propose Code
\ref{code:ancestor-permute-parallel}. This algorithm does not perform
the permutation in-place, but instead produces a new vector
$\mathbf{c} \in \{1,\ldots,N\}^N$ that is the permutation of the input
vector $\mathbf{a}$. It introduces a new vector $\mathbf{d} \in
\{1,\ldots,N + 1\}^N$, through which, ultimately, $c^i = a^{d^i}$. In
the first stage of the algorithm, \proc{Prepermute}, the thread for
element $i$ attempts to claim position $a^i$ in the output vector by
setting $d^{a^i} \leftarrow i$. By virtue of the $\min$ function on
line \ref{line:prepermute-min}, the element of lowest index always
succeeds in this claim while all others contesting the same place
fail, and the outcome of the whole permutation procedure is
deterministic. This is desirable so that the results of a particle
filter are reproducible for the same pseudorandom number seed. For
each element $i$ that is not successful in its claim, the thread for
$i$ instead attempts to claim $d^i$, if unsuccessful again then
$d^{d^i}$, then recursively $d^{d^{d^i}},\ldots$ etc, until an
unclaimed place is found.

\begin{code}[ht]
\begin{codebox}
\Procname{$\proc{Prepermute}(\mathbf{a} \in \{1,\ldots,N\}^N)
  \rightarrow \{1,\ldots,N + 1\}^N$}
\li Let $\mathbf{d} \in \{1,\ldots,N + 1\}^N$ and set $d^i \leftarrow N + 1$ for $i = 1,\ldots,N$.
\li \Foreach $i \in \{1,\ldots,N\}$
\li   \Atomic $d^{a^i} \leftarrow \func{min}(d^{a^i}, i)$ \label{line:prepermute-min} \Comment attempt to claim this slot, minimum is winner
    \End
\li \Ensures
\li   $\forall i (i \in \{1,\ldots,N\}: o^i > 0 \implies d^i = \min_j (a^j = i))$
    \End
\li \Return $\mathbf{d}$
\end{codebox}

\begin{codebox}
\Procname{$\proc{Permute}(\mathbf{a} \in \{1,\ldots,N\}^N)
  \rightarrow \{1,\ldots,N\}^N$}
\li $\mathbf{d} \leftarrow \proc{Prepermute}(\mathbf{a})$
\li \Foreach $i \in \{1,\ldots,N\}$
\li   $x \leftarrow d^{a^i}$
\li   \If $x \neq i$ \Comment if claim was unsuccessful in \proc{Prepermute} \Then \label{line:ancestor-permute-if}
\li     $x \leftarrow i$
\li     \While $d^x \leq N$ \label{line:ancestor-permute-while}
\li       $x \leftarrow d^x$
        \End
\li     $d^x \leftarrow i$
      \End
    \End
\zi
\li \Foreach $i \in \{1,\ldots,N\}$
\li   $c^{i} \leftarrow a^{d^i}$
    \End
\zi
\li \Ensures
\li    $\forall i (i \in \{1,\ldots,N\}: o^i > 0 \implies c^i = i)$
    \End

\li \Return $\mathbf{c}$
\end{codebox}

\caption{Parallel algorithm for the permutation of an ancestry vector to
    ensure condition \eqref{eqn:predicate} in the main article.}
\label{code:ancestor-permute-parallel}
\end{code}

We offer a proof of the termination of Code
\ref{code:ancestor-permute-parallel}. First note that \proc{Prepermute} leaves
$\mathbf{d}$ in a state where, excluding all values of $N + 1$, the remaining
values are unique. Furthermore, in \proc{Permute} the conditional on line
\ref{line:ancestor-permute-if} means that the loop on line
\ref{line:ancestor-permute-while} is only entered for values of $i$ that are
not represented in $\mathbf{d}$.

For each such $i$, the while loop traverses the sequence $x_0 = i$, $x_n =
d^{x_{n-1}}$, until $d^{x_n} = N + 1$. For the procedure to terminate this
sequence must be finite. Because each $x_n$ is an element of the finite set
$\{1,\ldots,N\}$, to show that the sequence is finite it is sufficient to show
that it never revisits the same value twice. The proof is by induction.
\begin{proof}
\begin{enumerate}
\item As no value of $\mathbf{d}$ is $i$, the sequence cannot revisit its
  initial value $\mathbf{x}_0 = i$. The element $x_0$ is therefore unique.
\item For $k \geq 1$, assume that the elements of $x_{0:k-1}$ are unique.
\item Now, the elements of $x_{0:k}$ are \textsl{not} unique if there exists
  some $j \in \{1,\ldots,k-1\}$ such that $x_k = d^{x_{k-1}} = x_j =
  d^{x_{j-1}}$, with $x_{j-1} \neq x_{k-1}$ by the uniqueness of
  $x_{0:k-1}$. But this contradicts the uniqueness of the (non $N + 1$) values
  of $\mathbf{d}$. Thus the elements of $x_{0:k}$ are unique, the sequence is
  finite, and the program must terminate.
\end{enumerate}
\end{proof}

\section{Auxiliary functions}\label{sec:auxiliary}

The multinomial, Metropolis and rejection resamplers most naturally
return the ancestry vector $\mathbf{a}$, while the stratified and
systematic resamplers return the cumulative offspring vector
$\mathbf{O}$. Conversion between these is reasonably
straightforward. An offspring vector $\mathbf{o}$ may be converted to
a cumulative offspring vector $\mathbf{O}$ via the
\proc{Inclusive-Prefix-Sum} primitive, and back again via
\proc{Adjacent-Difference}. A cumulative offspring vector may be
converted to an ancestry vector via Code
\ref{code:cumulative-offspring-to-ancestors}, and an ancestry vector
to an offspring vector via Code
\ref{code:ancestors-to-offspring}. These functions perform well on
both CPU and GPU. An alternative approach to
\proc{Cumulative-Offspring-To-Ancestors}, using a binary search for
each ancestor, was found to be slower.

\begin{code}[ht]
\begin{codebox}
\Procname{$\proc{Cumulative-Offspring-to-Ancestors}(\mathbf{O} \in \{0,\ldots,N\}^N) \rightarrow \{1,\ldots,N\}^N$}
\li \Foreach $i \in \{1,\ldots,N\}$
\li   \If $i = 1$ \Then
\li     $\id{start} \leftarrow 0$
\li   \Else
\li     $\id{start} \leftarrow O^{i-1}$
      \End
\zi
\li   $o^i \leftarrow O^i - \id{start}$
\li   \For $j = 1,\ldots,o^i$
\li     $a^{\id{start} + j} \leftarrow i$
      \End
    \End
\li \Return $\mathbf{a}$
\end{codebox}

\caption{Pseudocode conversion of an offspring vector $\mathbf{o}$ to an
  ancestry vector $\mathbf{a}$.}
\label{code:cumulative-offspring-to-ancestors}
\end{code}

\begin{code}[ht]
\begin{codebox}
\Procname{$\proc{Ancestors-to-Offspring}(\mathbf{a} \in \{1,\ldots,N\}^N)
  \rightarrow \{0,\ldots,N\}^N$}
\li $\mathbf{o} \leftarrow \mathbf{0}$
\li \Foreach $i \in \{1,\ldots,N\}$
\li   \Atomic $o^{a^i} \leftarrow o^{a^i} + 1$
    \End
\li \Return $\mathbf{o}$
\end{codebox}

\caption{Pseudocode conversion of an ancestor vector $\mathbf{a}$ to an
  offspring vector $\mathbf{o}$.}
\label{code:ancestors-to-offspring}
\end{code}

\section{Implementation}\label{sec:implementation}

All the algorithms described in the article and the appendices have been implemented as part of the
LibBi software (\url{www.libbi.org}, \citet{Murray2013b}) for performing
methods such as the particle filter on high-performance computing devices. We
enumerate the most important considerations of the implementation here, and
avoid painstaking detail of the remainder so as not to oversell their
importance relative to these. It is worth emphasising that some important
decisions, such as the choice of pseudorandom number generator (PRNG), depend
on the particular problem at hand.

The weight vector may contain many very small values. Because of this, a
typical implementation will store \emph{log-weights} rather than
\emph{weights} for numerical accuracy. The log-weights may be large and
negative, and one should avoid taking a floating point exponential of these
large negative numbers, which is often zero. All of the algorithms presented
are robust to the scaling of weights by a constant factor, however. When
computing sums or prefix sums, a vector of log-weights can therefore be
renormalised using, say, the maximum value, denoted $\log w_{\text{max}}$. For example,
the logarithm of the sum of weights, stored as log-weights, is accurately
computed using the identity:
\begin{equation*}
\log \sum_{i=1}^N w^i = \log w_{\text{max}} + \log \sum_{i=1}^N \exp (\log w^i - \log
w_{\text{max}}).
\end{equation*}
Renormalisation is not required for the Metropolis and rejection
algorithms, as they feature only pairwise ratios between weights, or pairwise
differences between log-weights.

The performance of the multinomial, stratified and systematic resamplers
depends largely on the implementation of the prefix sum operation. We defer to
existing work for these operations, in particular to that invested in the
Thrust library~\citep{Bell2012}, which the implementation uses. Conceptually,
the implementation in the Thrust library is based on up- and down-sweeps of a
balanced binary tree~\citep{Harris2007}.

The performance of the Metropolis and rejection resamplers is dependent mostly
on the selection of PRNG. Performance is not the only consideration in this
selection, however. PRNGs are assessed both on execution speed and the
statistical quality of the pseudorandom number sequence that they produce,
typically using test suites such as DIEHARD~\citep{Marsaglia1996} or
TestU01~\citep{LEcuyer2007}. Among clients of PRNGs, Monte Carlo algorithms,
such as the particle filter, have high demands for statistical quality. To
this end, our CPU code uses the Mersenne Twister PRNG~\citep{Matsumoto1998a}
as implemented in the Boost.Random library (\url{www.boost.org}). This is
standard for Monte Carlo applications. Our GPU code uses the XORWOW
PRNG~\citep{Marsaglia2003} from the CURAND library~\citep{CURAND}. This
particular PRNG belongs to a family that is readily shaped to the GPU
architecture~\citep{Nandapalan2012}. Faster but lower quality PRNG may be
used. This would constitute a relaxing of the unbiasedness condition
(2). As any such decision is problem-specific, it is not
investigated in this work.

The Metropolis and rejection algorithms use random access patterns to
memory. Spatiotemporally local access patterns are preferred for good
cache performance on CPU, and streaming, or at least coalesced access,
is preferred on GPU. The random access pattern is, unfortunately,
inherent to the algorithms, and we can only rely on the presence of a
large cache to mitigate associated latencies. On GPU, juditious use of
shared memory may help, but there is no reason to believe that this
can achieve better results than the hardware-controlled cache found on
more recent architectures; we rely on the latter. As such, the GPU is
configured to use 48 KB of L1 cache and 16 KB of shared memory. This
maximises the size of the cache for random access patterns, but still
provides sufficient shared memory for all kernels.

Our Metropolis and rejection resampler kernels compile to 32 registers per
thread, as reported by the CUDA compiler. This is satisfactory with respect to
occupancy of the device, and we do not seek further reductions.

Finally, the auxiliary algorithms presented in \S\ref{sec:auxiliary} pose
little challenge. Implemented using CUDA, they compile to kernels using no
shared memory and fewer than 16 registers per thread, which is of no hindrance
to occupancy of the device. On GPU, we append the \proc{Prepermute} procedure
of Code \ref{code:ancestor-permute-parallel} to the end of any procedure that
produces an ancestry vector. This saves the launch of a separate kernel and
the associated overhead of doing so.

\end{document}